\documentclass[
aps,
prl,
12pt,
reprint,
longbibliography,
floatfix
]{revtex4-2}

\usepackage[utf8]{inputenc}
\usepackage[T1]{fontenc}
\usepackage[english]{babel}
\usepackage{amsmath}
\usepackage{physics}
\usepackage{graphicx}
\usepackage[colorlinks=true, allcolors=black]{hyperref}
\usepackage{xcolor}
\usepackage{graphicx}
\usepackage[left=23mm,right=13mm,top=35mm,columnsep=15pt]{geometry} 
\usepackage{placeins}
\usepackage[T1]{fontenc}
\usepackage{appendix}

\usepackage{comment}
\usepackage{color}

\graphicspath{{Images/}}

\begin{document}

\title{Phenomenology of a Rydberg impurity in an ideal Bose Einstein condensate }
\author{Aileen A. T. Durst}
\email{dursta@pks.mpg.de}
\author{Matthew T. Eiles}
\email{meiles@pks.mpg.de}
\affiliation{Max Planck Institute for the Physics of Complex Systems,  Nöthnitzer Str. 38, 01187 Dresden, Germany}

\date{\today} 

\begin{abstract}

We investigate the absorption spectrum of a Rydberg impurity immersed in and interacting with an ideal Bose-Einstein condensate.
Here, the impurity-bath interaction can greatly exceed the mean interparticle distance; this discrepancy in length scales challenges the assumptions underlying the  universal aspects of impurity atoms in dilute bosonic environments.
Our analysis finds three distinct parameter regimes, each characterized by a unique spectral response. 
In the low-density regime, we find that the Rydberg impurity is dressed by the surrounding bath similarly to the known Bose polaron. 
Transitioning to intermediate densities, the impurity response, given by sharp quasiparticle peaks, fragments into an intricate pattern bearing the hallmarks of a diverse molecular structure.
Finally, at high density, a universal Gaussian response emerges as the statistical nature of the bath dominates its quantum dynamics. 
We complement this analysis with a study of an ionic impurity, which behaves equivalently. 
Our exploration offers insights into the interplay between interaction range, density, and many-body behavior in impurity systems. 

\end{abstract}

\maketitle
The dynamics of strongly-correlated quantum mixtures pose a significant challenge to theoretical description, even at the level of a single impurity immersed in a non-interacting bath. 
The apparent simplicity of the Hamiltonian of such a mixture,
\begin{align}\label{eq:H}
	\hat H &= \sum_{\pmb{k}} \frac{\pmb{k}^2}{2\mu}\hat b_{\pmb{k}}^\dagger \hat b_{\pmb{k}} + \sum_{\pmb{k},\pmb{q}} V(\pmb{q})  \hat b_{\pmb{k}+\pmb{q}}^\dagger \hat b_{\pmb{k}},
\end{align}
belies the rich complexity of the phenomena emergent in this many-particle system \cite{schmidt_universal_2018, alexandrov_advances_2009, shchadilova_quantum_2016}. 
In \autoref{eq:H}, which is written in a frame centered on the zero-momentum impurity,  $\hat b_{\pmb{k}}^\dagger$ and $\hat b_{\pmb{k}}$ denote the bath creation and annihilation operators, $V(\pmb{q})$ is the interspecies interaction, and $\mu$ is the reduced mass of the impurity and a bath atom \cite{lee_interaction_1953}.  
$\hat H$ is commonly used to describe dilute gases with a mean interparticle distance $\rho^{-1/3}$, where $\rho$ is the density, greatly exceeds the range of the potential $V(\pmb{r})$. 
This justifies its replacement by a zero-range pseudopotential proportional to the bath-impurity scattering length $a$ \cite{huang1957quantum}. 
Within this approximation, the physics of the system becomes universal, depending only on the scattering length, the dimensionality of the system, and the density and quantum statistics of the bath \cite{shashi_radio-frequency_2014, rath_field-theoretical_2013, christensen_quasiparticle_2015, drescher_real-space_2019, ardila_dynamical_2021, van_loon_ground-state_2018, grusdt_strong-coupling_2017, tajima_polaron_2021, massignan_polarons_2014,chevy_ultra-cold_2010,etrych2024universal}.
Measurements of repulsive and attractive polaron quasiparticles and weakly bound molecules in ultracold gases have provided strong evidence for this universal behavior \cite{jorgensen_observation_2016, yan_bose_2020, hu_bose_2016, catani_quantum_2012, schmidt_quantum_2018, skou_non-equilibrium_2021,scelle_motional_2013, ness_observation_2020, schirotzek_observation_2009,chin_feshbach_2010,etrych2024universal}.

\begin{figure}
\centering
\includegraphics[width = 0.5 \textwidth]{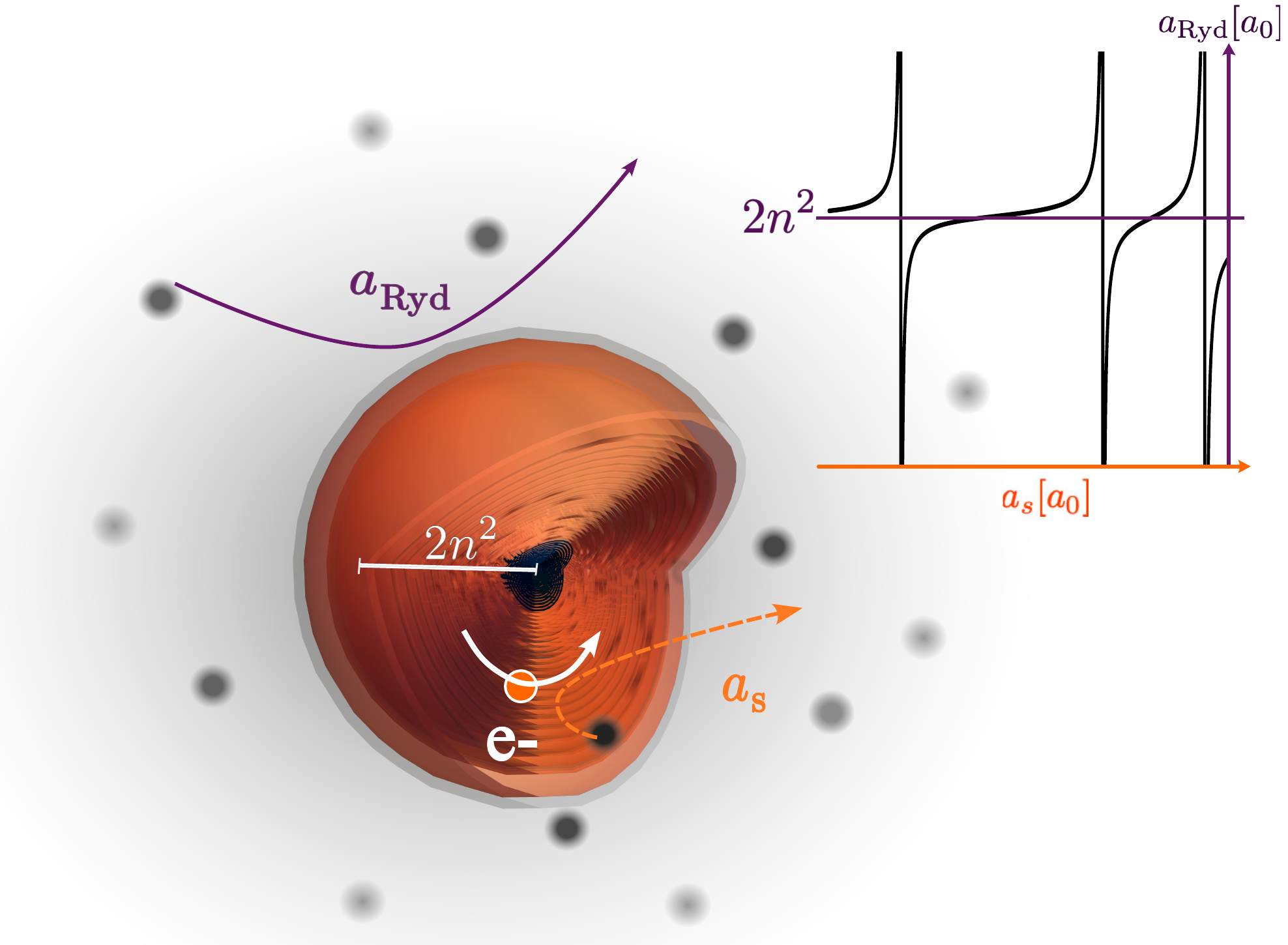}
\caption{Two scattering lengths characterize the interaction of a Rydberg impurity with its environment. Atoms probing the interior of the Rydberg atom collide with the highly excited electron directly; these interactions are characterized by the atom-electron scattering length $a_\mathrm{s}$. In contrast, distant atoms interact with the Rydberg atom as a single entity, and the atom-impurity scattering length $a_\mathrm{Ryd}$ is characteristic of this interaction. }
\label{fig1}
\end{figure}

However, this universality is not expected to apply when the interaction range is comparable to the typical interparticle distance.
Such is the case for a Rydberg impurity, where the highly excited Rydberg electron with principal quantum number $n$ mediates the impurity-bath interaction by scattering off of the bath particles. 
The $s$-wave electron-atom scattering length $a_s$ (see \autoref{fig1}) determines the overall strength of this interaction \cite{greene_creation_2000,fermi_sopra_1934, omont_theory_1977}.
For a Rydberg $\ket{nS}$ state this leads to the isotropic potential 
\begin{align}\label{eq:Ryd_pot}
V_\mathrm{Ryd} (r) = 2\pi a_s \abs{\psi_{n00}(r)}^2,
\end{align}
which, unlike a zero-range potential, can in principle support several bound states \cite{bendkowsky_observation_2009, shaffer_ultracold_2018, fey_ultralong-range_2020, sasmannshausen_long-range_2016,eiles_trilobites_2019}. 
The appearance of the Rydberg wave function, $\psi_{n00}(r)$, causes the range $R_0$ and depth $V_0$ of this highly oscillatory potential to vary as $n^2$ and $n^{-6}$, respectively \cite{eiles_trilobites_2019}.
Recently, the ultracold toolbox has been expanded to include other impurity systems with finite-ranged interactions, such as dipolar atoms \cite{volosniev_non-equilibrium_2023,aikawa_bose-einstein_2012,lahaye_strong_2007} and ion-atom mixtures \cite{christensen_charged_2021, astrakharchik_ionic_2021, pessoa_fermi_2024, massignan_static_2005}.
These break the zero-range universality in a similar fashion, and raise the question of whether or not a unified description of a finite-ranged impurity in a quantum environment exists. 

In this article, we approach this question through an exploration of the behavior of a Rydberg impurity interacting with an ideal Bose-Einstein Condensate (BEC) at zero temperature. 
Previous studies \cite{schmidt_mesoscopic_2016, camargo_creation_2018, schmidt_theory_2018, sous_rydberg_2020} have treated such an impurity in isolation from the zero-range Bose polaron due to the large discrepancy in length and energy scales. 
However, we show that these different impurities share the same underlying physics determined by the universal parameter $a\rho^{1/3}$. 
Even though the spectral response becomes more complicated in finite-ranged impurity systems, each component can be understood and generally described. 
To further support these findings, we also consider an ionic impurity. 
Together, this leads to an extension of the universal description of Bose polarons to a broader class of interactions. 
Further, we attempt to unify the disparate interpretations provided by the many approaches developed for such impurity problems, which include the field-theoretical quasiparticle methods describing polaron physics, the few-body picture of molecular physics, and semiclassical methods originating in the theory of pressure broadening.

To investigate and characterize the universal aspects of the Rydberg impurity, we performed a detailed numerical study of the absorption spectrum $A(\omega)$. 
This is obtained from the Fourier transform of the auto-correlation function
\begin{equation}\label{Eq:LSE}
S(t) = \langle e^{i \hat H_0t}e^{-i \hat H t}\rangle = \left(\sum_\alpha e^{i(\epsilon_0-\omega_\alpha) t } \abs{ \bra{0}\ket{\alpha}}^2\right)^N,
\end{equation}
where the expectation value is taken with respect to the non-interacting BEC state (the ground state of $\hat H_0$) \cite{schmidt_theory_2018}. Since the ideal BEC is in a pure product state, the final expression of the many-body response only requires the eigenstates $(\ket 0)$, $\ket{\alpha}$ and energies $(\epsilon_0)$, $\omega_\alpha$ of the (non)-interacting two-body Hamiltonian of the Rydberg atom and a single boson.  
We solve for the two-body physics by employing the eigenchannel R-matrix method \cite{aymar_multichannel_1996}, which yields the energy-dependent logarithmic derivative of the scattering wave function $\ket{\alpha}$ for $r>R_0$. 
This allows us to efficiently calculate thousands of box-continuum states, molecular bound states, and the zero-energy scattering length $a_\mathrm{Ryd}$. 
\autoref{fig1} shows $a_\mathrm{Ryd}(a_s)$, which carries information about both the interaction range $R_0\sim 2n^2$ and the dependence of two-body bound states on $V_0$. 
We study a Rydberg impurity with $n=50$ as an illustrative and generic example, and compute $A(\omega)$ as a function of $a_s$ \footnote{Further details about the interaction potential and the numerical parameters for the calculation can be found in the Supplementary Material at: URL to be inserted by publisher.}.
In this way we adjust $V_0$ independent of $R_0$ \footnote{While it is not possible to tune the electron-atom scattering length in experiment, it is a convenient theoretical tool due to this decoupling of $V_0$ and $R_0$. To change $a_\mathrm{Ryd}$ experimentally, $n$ can be varied to explore the different interaction regimes studied here.}.

\begin{figure}[b]
	\includegraphics[width=1 \columnwidth]{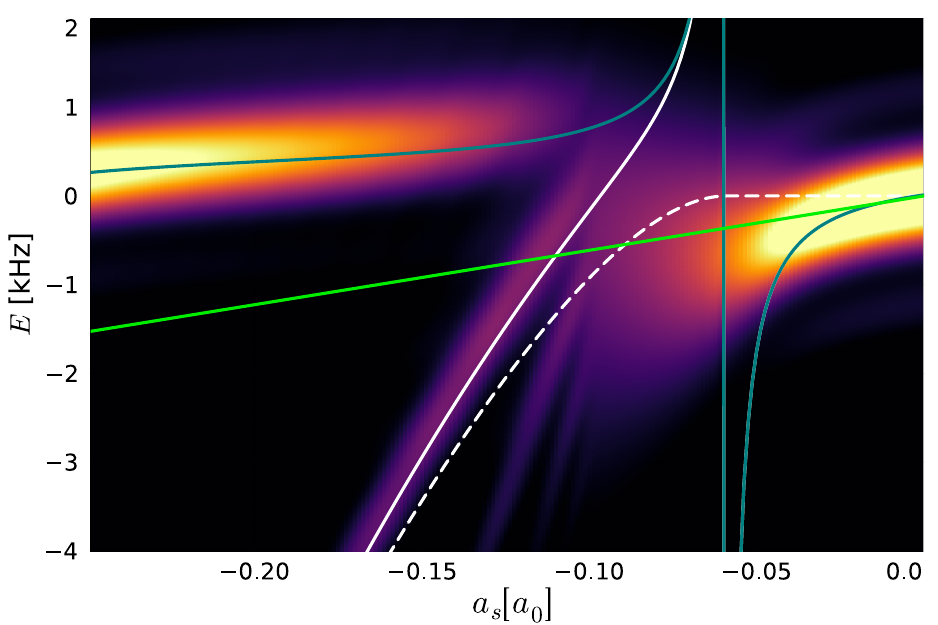}
	\caption{$A(\omega)$ of a $50S$ Rydberg impurity in a BEC with $\rho=10^{12}\,$cm$^{-3}$. The mean-field energy shifts $E_\mathrm{zr}$ (teal) and $E_\mathrm{Ryd}$ (green), as well as the bare and dressed dimer energies $E_\mathrm{b}$ (solid white) and $E_\mathrm{b}+E_\mathrm{zr}$ (dashed white) are overlaid.}
	\label{fig:fig2_mte}
\end{figure}

\autoref{fig:fig2_mte} shows $A(\omega)$ when $R_0\rho^{1/3}\ll 1$.
 In this regime, we recover all of the features known from the limit of a zero-range impurity \cite{drescher_quench_2021, shchadilova_quantum_2016, grusdt_strong-coupling_2017}.
When $a_\mathrm{s}>-0.06a_0$, $V_\mathrm{Ryd}$ does not support a two-body bound state, and $A(\omega)$ exhibits a single peak at negative energy indicating the formation of an attractive polaron. 
The mean-field energy of the zero-range approximation of the Rydberg potential, $E_\mathrm{zr} = 2\pi a_\mathrm{Ryd}\rho/\mu$, describes the position of this feature.
However, this description fails dramatically near a scattering resonance.
Instead, the mean-field description of the full Rydberg potential, 
$E_\mathrm{Ryd} =\rho \int V_\mathrm{Ryd}(r)\mathrm{d}^3r = {2\pi a_s\rho}/{m_e}$, \cite{schmidt_theory_2018, fermi_sopra_1934, allard_effect_1982} follows the center of spectral weight smoothly across unitarity, even as the spectral feature diffuses and cannot be associated with a well-defined quasiparticle  \cite{etrych_universal_2024, shchadilova_quantum_2016}. 
Using the Born approximation for the Rydberg scattering length, $a_\mathrm{Ryd}^\mathrm{B}=a_s\mu/m_e$, one can rewrite $E_\mathrm{Ryd}=2\pi a_\mathrm{Ryd}^\mathrm{B}\rho/\mu$ to have the same structure as $E_\mathrm{zr}$.

To the red of the resonance, $E_\mathrm{zr}$ again describes the brightest spectral feature, which is located at positive energy and therefore identified as a repulsive polaron. 
Below this feature lies a series of negative energy peaks associated with the molecular bound state, which can be multiply occupied to form dimers, trimers, and the like. 
These interact with the residual bath through the same Rydberg interaction as the bare atom, and as a result they are dressed by bath excitations in the same fashion. 
Each ultralong-range Rydberg molecule therefore possesses some quasiparticle character inherited from the repulsive polaron and forms a "molaron" with a binding energy shifted from that of the bare dimer, $E_\mathrm{b}$, to  $E_\mathrm{b} + E_\mathrm{zr}$ 
\cite{shchadilova_polaronic_2016, mostaan_unified_2023, diessel_probing_2022, Schirotzek_thesis}. 
Myriad experiments have observed this vibrational spectrum, confirming its basic structure but not yet providing conclusive evidence for the many-body shift $E_\mathrm{zr}$ \cite{bendkowsky_rydberg_2010,gaj_molecular_2014,engel_precision_2019,camargo_creation_2018}.
This is not surprising, since theoretical uncertainties \cite{fey_comparative_2015,greene_greens-function_2023,maclennan_deeply_2019,peper_photodissociation_2020,giannakeas_generalized_2020} would have hidden this small shift.

\begin{figure}
	\includegraphics[width=0.9 \columnwidth]{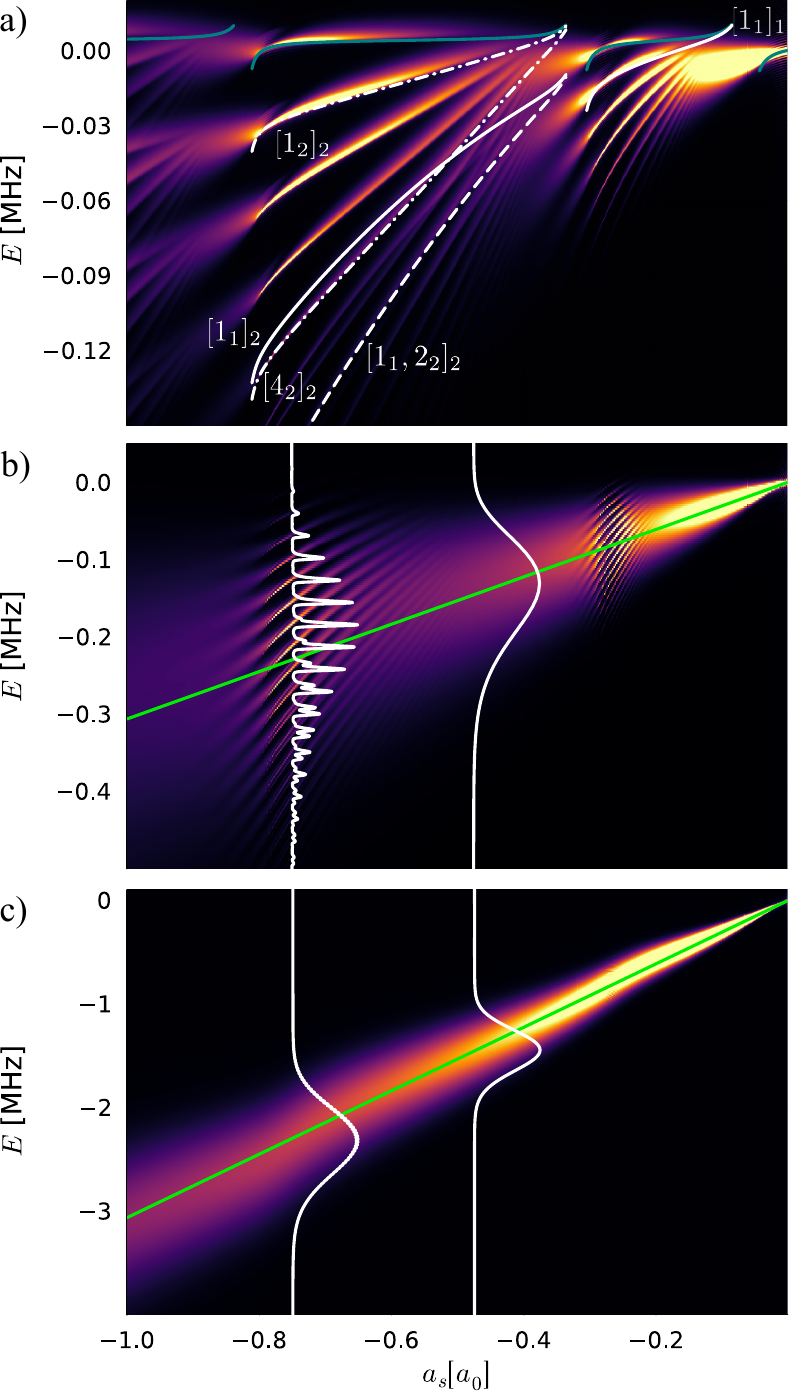}
	\caption{$A(\omega)$ of a $50S$ Rydberg impurity in a BEC with (a) $\rho=10^{13}\,$cm$^{-3}$,  (b) $\rho=5 \cdot 10^{13}\,$cm$^{-3}$, and (c) $\rho=5 \cdot 10^{14}\,$cm$^{-3}$. In (a), the mean-field energy  $E_\mathrm{zr}$ (teal) is shown alongside the mean-field energies of various molaron states. 
 In panels (b) and (c) cuts of $A(\omega)$ at fixed $a_s$ (white curves) show the lineshape more clearly. $E_\mathrm{Ryd}$ is shown in green. }
	\label{fig:fig3}
\end{figure}

\autoref{fig:fig3}(a) displays $A(\omega)$ over a broader range of $a_s$ and at ten times the density of \autoref{fig:fig2_mte}, which causes the molaron peaks to accumulate appreciable spectral weight as $R_0\rho^{1/3}\sim 1$. 
This reveals an intriguing internal structure due to the appearance of additional two-body bound states as the potential deepens.
This structure suggests a nomenclature, the $[n_0,\dots, n_i,\dots,n_M]_M$-molaron, where $M$ is the total number of two-body bound states supported by the potential and $n_i$ is the bosonic occupation number of the $i$th bound state, with $i=0$ representing the bare atom. 
The peak position of the $[0_0]_M$-molaron, i.e. the polaron, coincides with $E_{zr}$.
Exemplary molaron peaks are labeled in \autoref{fig:fig3}(a).  
At every scattering resonance each quasiparticle peak undergoes the same fragmentation seen in \autoref{fig:fig2_mte}.
For example, at the second scattering resonance, each of the $[n_i]_1$ states broadens and eventually splits into a multitude of states $[n_i,m_{i+1}]_{2}$. 

An important consequence of the large extent of the Rydberg potential is that  $a_\mathrm{Ryd}$ is large and positive except close to a resonance, where it dips below zero.
As a result, the attractive polaron exists only in a very limited parameter space. 
At the transition from a repulsive to an attractive polaron, $a_\mathrm{Ryd}$ vanishes at a Ramsaeur-Townsend zero \cite{ramsauer_uber_1921, townsend_xcvii_1921}. 
Despite its non-zero interaction potential, the Rydberg atom effectively does not interact with the bath -- the scattering phase shift vanishes.
Here, the molaron features sharpen, losing quasiparticle weight as they more closely resemble bare molecules. 

At higher density (\autoref{fig:fig3}b) the spectral weight shifts entirely into molaron states.  
Their absorption peaks blur together and $A(\omega)$ takes on a Gaussian profile with mean values $E_\mathrm{Ryd}$ whose emergence is explained by the central limit theorem \cite{schmidt_mesoscopic_2016,schmidt_theory_2018,royer_shift_1980}.
However, the regression to a Gaussian distribution does not occur at the same density for each $a_s$: the distinct molaron peaks, still resolvable in the vicinity of the Ramsauer-Townsend zeros in \autoref{fig:fig3}(b), merge to form a Gaussian spectral profile only at higher density (panel (c)).
\begin{figure}[t]
	\includegraphics[width=1 \columnwidth]{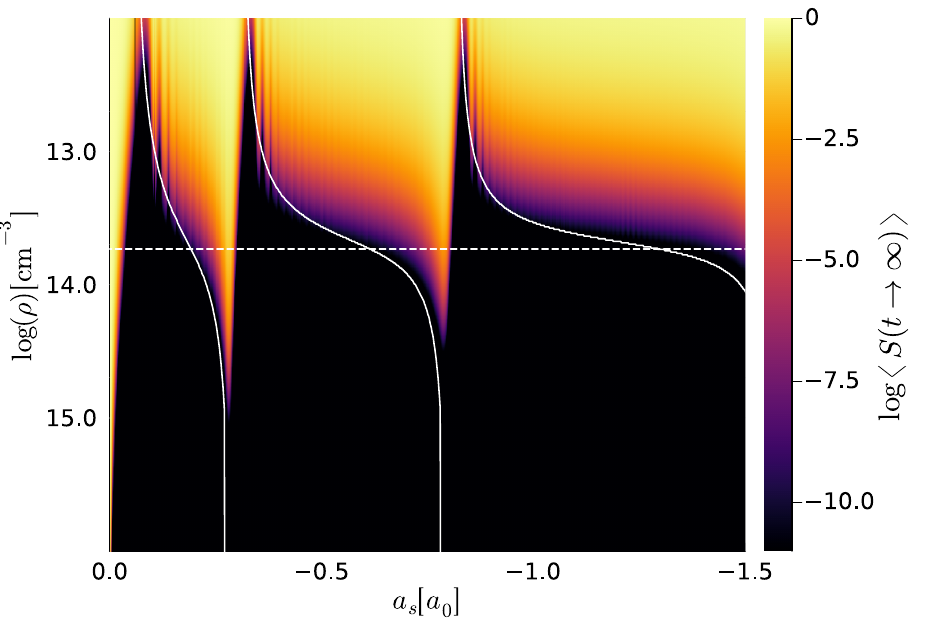}
	\caption{Effective quasiparticle weight of the Rydberg impurity, calculated by averaging $S(t)$ over late times. The solid white curve shows $a_\mathrm{Ryd}\rho^{1/3}=1$, which demarcates the two phases: quasiparticle (molaron or polaron limit with $Z=1$) and semiclassical / statistical ($Z=0$). The dashed white line shows $R_0\rho^{1/3}=1$, the semiclassical transition. }
 \label{fig:phaseportrait}
\end{figure}

This progression from an individual Lorentzian at zero density through asymmetric lineshapes and polymer formation to a symmetric Gaussian distribution at high density is largely consistent with the semi-classical theory of pressure broadening \cite{baranger_general_1958,baranger_simplified_1958, allard_collision-broadened_1988, allard_effect_1982, allard_alkali-rare-gas_1978, allard_alkali_1980, szudy_unified_1975, szudy_profiles_1996}. 
This theory predicts the occurrence of "satellite" peaks at integer multiples of the extrema of the interaction potential, an obvious parallel to the molaron structure, as well as their blending together into a Gaussian response as the number of particles within the range of the potential, $R_0^3\rho$, increases \cite{royer_shift_1980, allard_collision-broadened_1988}.
Our calculation shows that a more accurate condition takes into account the zero-energy scattering length, and thus generalizes the above condition to $\rho^{-1/3}\ll a_\mathrm{Ryd}$. 

This more accurate condition is depicted in \autoref{fig:phaseportrait}, which shows a qualitative estimate of the quasiparticle weight given by a temporal average of $S(t)$ in the long time limit. 
The curve $a_\mathrm{Ryd}\rho^{1/3}=1$ indicates the transition between a spectrum dominated by quasiparticle features (where $S(t)\gg 0$) to a purely statistical state characterized by the Gaussian lineshape with no quasiparticle weight.
In the extreme limit $a_\mathrm{Ryd}\to 0^{+/-}$, the above condition is never satisfied and $A(\omega)$ will show distinct peaks no matter the density.  
This explains the regions with large quasiparticle weight even at high densities seen in \autoref{fig:phaseportrait} where the impurity has a small but non-zero quasiparticle character due to its vastly reduced coupling to the dense environment.

In contrast, the region of approximately zero quasiparticle weight extends to very low densities when $a_\mathrm{Ryd}$ diverges. 
This corresponds to the parameter regime where the loss of quasiparticle characteristics and a broadening of the well-defined polaron peak into a diffuse continuum is known from zero-range impurities as well \cite{shashi_radio-frequency_2014, shchadilova_polaronic_2016, tempere_feynman_2009}.
Consequently, the spectral broadening of the attractive polaron peak near resonance in the zero-range polaron and the emergence of the Gaussian feature in Rydberg polaron studies share the same underlying physical origin. \\

This collection of phenomena is not limited to a Rydberg atom, but is shared by all impurity systems: the quantitative differences between the absorption spectra of various impurities are a matter of degree rather than of kind.
This can be seen by comparing the spectrum of a neutral impurity \cite{shchadilova_polaronic_2016,drescher_quench_2021} or an ionic impurity \cite{christensen_charged_2021} with the present results. 
	\begin{figure}[b]
		\includegraphics[width=1 \columnwidth]{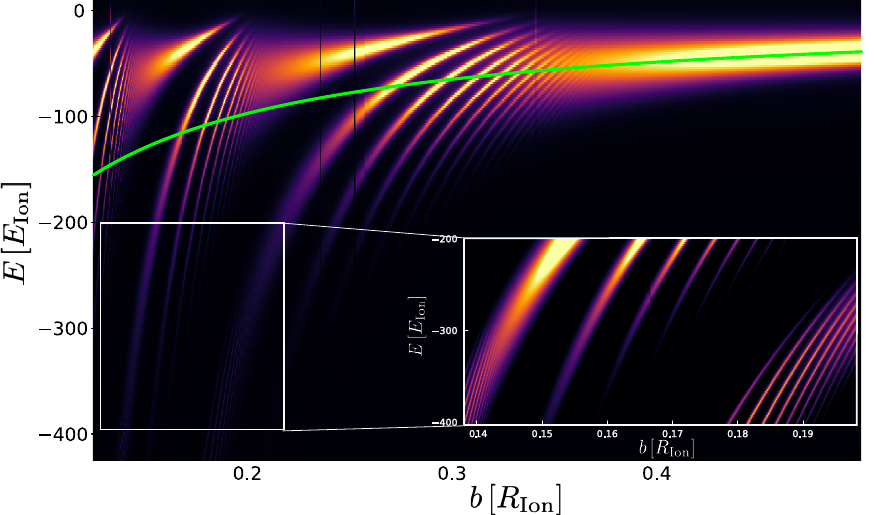}
		\caption{$A(\omega)$ of an ionic impurity at $\rho = 2\,R_\mathrm{Ion}^{-3}$. The green line shows the mean-field energy shift $E_\mathrm{Ion}$. The inset highlights the fragmentation of molaron states.}\label{fig:ion}
	\end{figure}
In the latter case,  the interaction is often taken to be a regularized polarization potential \cite{tomza_cold_2019,christensen_charged_2021}
	\begin{align}
		V_\mathrm{Ion}(r)&=-\frac{\alpha}{(r^2+b^2)^2}\frac{r^2-c^2}{r^2+c^2},
	\end{align}
with a characteristic length scale $R_\mathrm{Ion} =\sqrt{2\mu\alpha}$.
We calculated $A(\omega)$ for  $c=0.0023\,R_\mathrm{Ion}$, varying $b$ to adjust the potential depth, as shown in \autoref{fig:ion}. 
We observe polaron features and fragmentation of the molaron states as they cross a scattering resonance, familiar from \autoref{fig:fig3}(a). 
At the density of $\rho=2\,R_\mathrm{Ion}^{-3}$, the molaron states possess significant spectral weight and hint at the emerging Gaussian lineshape centered around the mean-field energy shift $E_\mathrm{Ion}=\rho \int V_\mathrm{Ion}(r)\mathrm{d}^3r$, which becomes completely dominant at higher densities as in \autoref{fig:fig3}(c) \cite{suppmat}. As with the Rydberg impurity, by writing $E_\mathrm{Ion} = 2\pi a_\mathrm{Ion}^{B}\rho/\mu$ we see that this feature is simply characterized by the Born approximation for the scattering length, the reduced mass and the bath density. \\

With the insights provided by these two impurity systems, we have shown that the universal parameter $a\rho^{1/3}$ determines the qualitative behavior of their absorption spectra. 
This unites the phenomena of finite-ranged impurities with those known from the well-studied zero-range impurity (the "Bose polaron"). 
Even deeply bound molecular states, whose energies depend on the details of the two-body interaction, respond identically to the influence of the bath.
At sufficiently high densities, these details become irrelevant and the system response is universal, depending only on its scattering length in the Born approximation within a mean-field description. 

The huge length scales of the Rydberg atom are especially appropriate for the approximations made in $\hat H$: the neglect of boson-boson interactions and kinetic energy correlation. 
Rydberg impurities therefore may serve as a more suitable platform for studying molarons and refining their theoretical description than ground-state atoms, even though the bath-induced density shift of the molecule peaks, typically on the scale $~4\pi n^2\rho/\mu$, is small relative to the binding energies. 
These subtle many-body energy shifts, including the shift of the bare atomic line (the repulsive Rydberg polaron), could be isolated by a careful study of the density-dependence and asymmetry of the absorption peaks, particularly for light atoms. 
Additionally, fermionic environments, which have been partially explored for Rydberg and ionic impurities, provide another avenue for future investigation \cite{sous_rydberg_2020, christensen_mobile_2022}.

\acknowledgements
We are grateful for many enlightening discussions with P. Giannakeas, A. Eisfeld, and S. Wüster.

\bibliography{MyLibrary}

\appendix
\section{Supplementary Information}
In the following, we provide additional details and parameters for the calculations presented in the main text. 
\subsection{Loschmidt Echo}
The Loschmidt-Echo $S(t)$, also known as the autocorrelation function, is the time-dependent overlap of the bare BEC state and the interacting state following the sudden introduction of the impurity at $t=0$. Beginning with the BEC wave function $\ket{\psi_\mathrm{BEC}} = 1/\sqrt{N!} b_0^{\dagger ^N}\ket{\mathrm{vac}}$, where 
\begin{align}
N = \frac{\rho_0 \cdot 2 L ^3}{\pi}
\end{align}
is the number of $s$-wave BEC particles, $S(t)$ is given by
\begin{align}\label{EQ:S_1}
S(t) &= \bra{\psi_\mathrm{BEC}}e^{i \hat H_0t}e^{-i \hat Ht}\ket{\psi_\mathrm{BEC}}\\
&= \left(\sum_\alpha e^{i(\epsilon_0-\omega_\alpha) t } \abs{ \bra{0}\ket{\alpha}}^2\right)^N.
\end{align}
Here the state $\ket 0$ is the ground state of the non-interacting two-body Hamiltonian 
\begin{equation}
    h_0(r) = -\frac{\nabla_r^2}{2\mu}
\end{equation}
with eigenenergy $\epsilon_0$, while $\ket{\alpha}$ are the eigenstates of the interacting two-body Hamiltonian
\begin{equation}\label{Eq:single_h}
h(r) = -\frac{\nabla^2_r}{2\mu}+V_\mathrm{Ryd}(r)
\end{equation}
with eigenenergies $\omega_\alpha$. 
Calculating $S(t$) ultimately involves determining the interacting eigenstates $\ket{\alpha}$ and their corresponding eigenenergies $\omega_\alpha$. A more complete derivation can be found in \cite{sous_rydberg_2020}. 
We perform the time evolution of $S(t)$ up to $t_\mathrm{max}= 1000\,\mu$s in order to compute $A(\omega)$ from directly integrating the Fourier transform of $S$. To have a numerically well-defined Fourier transformation we multiply $S(t)$ with a exponential decay $\exp[-t/ (0.4\cdot t_\mathrm{max})]$. 
This decay time is chosen to be large in the present study to avoid obscuring any interesting results by numerical broadening of the spectral features. 
Tests for shorter decay times, which model the finite Rydberg lifetime, did not show significant deviation from the calculations performed here.

\subsection{Rydberg potential}
For an $s$-state Rydberg electron the interaction potential $V_\mathrm{Ryd}(r)$ is isotropic and spherically symmetric. To simplify the theoretical analysis, we neglect the effect of a finite quantum defect and use the electronic wave function $\psi_{nlm}(r)$ of the hydrogen atom. Further, we assume that the electron-atom scattering length $a_s$, which sets the strength of the overall interaction, is energy-independent. 
These assumptions yield the two-body interaction potential,
\begin{align}
V_\mathrm{Ryd} (r) = 2\pi a_s \abs{\psi_{n00}(r)}^2,
\end{align}
which captures the features of more sophisticated calculations to a semi-quantitative degree. In the main text our results are computed for an electronic $n=50$ state of a mass-balanced system with the mass of $^{84}$Sr, $\mu = 83.91342/2 \,[\mathrm{au}]$,  in a box with radial extent of $L= 550 n^2 a_0$. 

Within the Born approximation, the zero-energy scattering length is given by $a_\mathrm{Ryd}^\mathrm{B}=\frac{m_e}{\mu}a_s$, from which we obtain the density shift $E_\mathrm{Ryd}= \frac{2 \pi a_s}{m_e}$. This gives the position of the Gaussian feature in the high density regime. 

\subsection{Ion potential}
The interaction between an ion and a neutral atom has a long-range tail $\propto -\alpha/(2r^4)$, with $\alpha$ the polarizability of the neutral atom. To avoid problems at short inter-particle distances we include a short-range regularization, which gives us the ionic interaction potential
\begin{align}
V(r)&=-\frac{\alpha/2}{(r^2+b^2)^2}\frac{r^2-c^2}{r^2+c^2}
\end{align}
with characteristic range $R_\mathrm{Ion}=\sqrt{2\mu\alpha}$. 
We fix the free parameters to be $c= 0.0023\, R_\mathrm{Ion}$, $\alpha = 320$, and $\mu = 86.9092/2\, [\mathrm{au}]$, corresponding to the polarizability and mass of $^{87}$Rb atoms.  

Within the Born approximation, the zero-energy scattering length is 
\begin{align}
a_\mathrm{Ion}^\mathrm{B} &= -\frac{R_\mathrm{ion}^2\pi}{4b}\frac{(b^2+2bc-c^2)}{(b+c)^2}.
\end{align}
From this, or alternatively from the integral $\rho\int V_\mathrm{Ion}(r)\mathrm{d}^3r$, we obtain the density shift $E_\mathrm{Ion}= \frac{2 \pi }{\mu} \rho a_\mathrm{Ion}^\mathrm{B}$ giving the position of the Gaussian feature in the high density regime. 

\subsection{Spectrum of $h$}
In the following, we describe our general approach to calculating the spectrum of a given two-body Hamiltonian $h$ with a generic interaction potential which can be truncated at some finite distance. 
When we give details, such as the number of basis states and dimensions of the quantization volume, they are specific to the calculation of the Rydberg impurity. The ionic impurity can require, due to the slower decay of its interaction, a larger basis size and matching of interacting and free wave fucntions at a larger radius to achieve convergence.

We separate the bound state calculation from the continuum state calculation in order to save computational effort, since we want to avoid diagonalizing a huge matrix to obtain many hundreds of continuum states. 

\begin{figure}
\centering
		\includegraphics[width=0.4\textwidth]{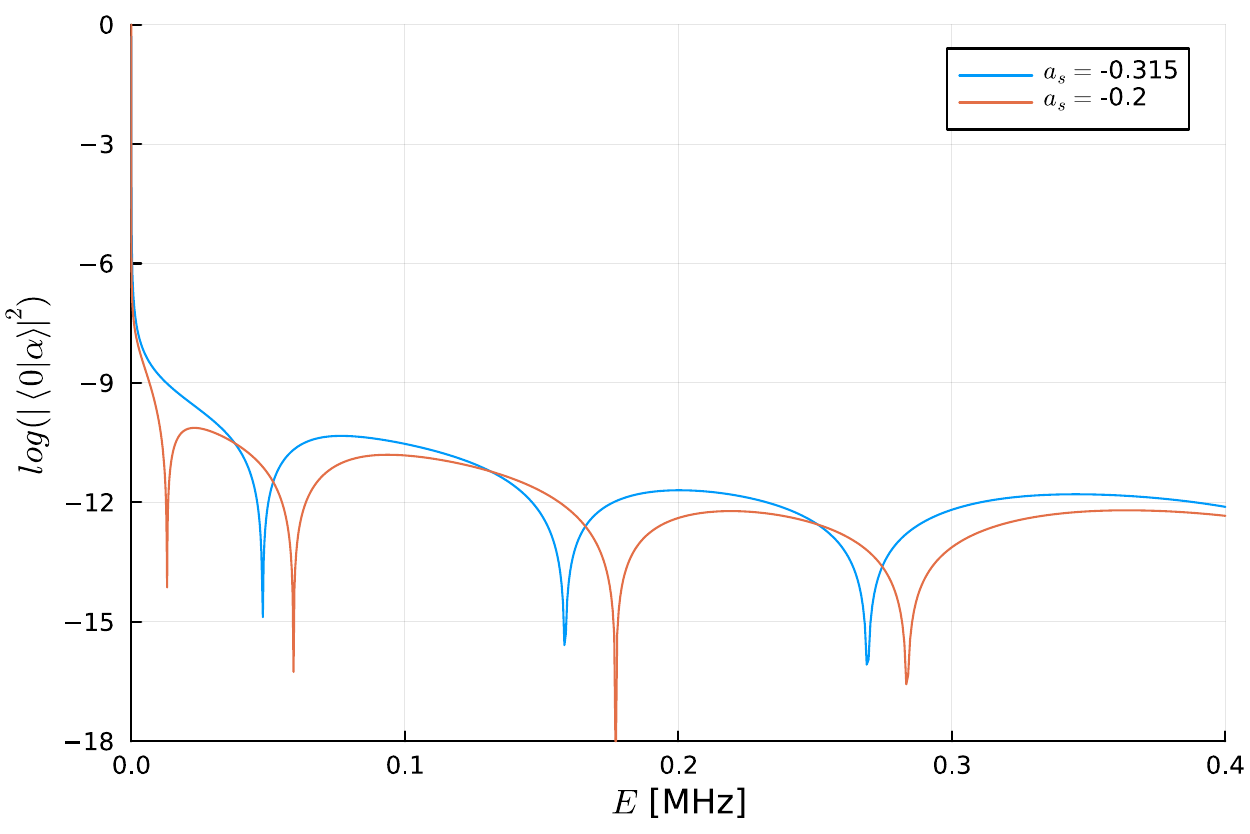}
\caption{The energy-dependent Frank-Condon overlaps of the continuum states of $h$ with the ground state of the non-interaction system $h_0$.}
\label{fig:Overlap}
\end{figure}

In our calculations for the continuum of single-particle states $\ket{\alpha}$, we employ the eigenchannel R-matrix approach.  This method obviates the need to solve the Schrödinger equation numerically over the entire quantization volume. We partition the space around the impurity into two distinct regions: one (roughly for $0<r<3n^2$) where the Rydberg potential differs significantly from zero, and another $3n^2<r<L$ where the interaction potential is negligible and the wave function is known analytically.
At the boundary of the interaction volume, we compute the log-derivative of the wave function at a specific energy $E$,
\begin{equation}
-b_\beta(E) = \frac{\partial \ln r \psi_\beta}{\partial r} .
\end{equation}
We compute $b_\beta$ using the streamlined eigenchannel approach detailed in Ref.~\cite{aymar_multichannel_1996}, using 500 B-spline functions of order 12 to span the range from the inner boundary $r_0 =200 a_0$ to $r_1 = 3 n^2$. 
By now matching the analytical log-derivative of the free particle solutions outside of the range of the interaction potential with those inside, we compute the energy-dependent phase shift $\delta(E)$ and the scattering length $a_\mathrm{imp}$ of the potential as follows:
\begin{align}
\tan(\delta(E)) &= \frac{b(E) j_0(k r_1)r_1 + \partial_r j_0(k r_1)}{b(E) y_0(k r_1)r_1 + \partial_r y_0(k r_1)}\\
a_\mathrm{imp} &= - \mathrm{lim}_{k \rightarrow 0 }\tan(\delta(k))/k.
\end{align}
In a second step we discretize the continuum by imposing the hard wall boundary condition at $r = L$. To achieve this, we do an energy search for all wave functions that have zero amplitude at the box boundary. 
In total we use about 10000 interacting states $\ket{\alpha}$ up to a energy cut-off ($E_\mathrm{max} = 300\,$[MHz]) to represent the continuum. Especially close to a resonance the continuum couples strongly and a good numerical representation of the continuum becomes particularly important.
The calculated overlaps of the low-lying box-continuum states with the free BEC wave function is shown in \autoref{fig:Overlap}, for two different interaction strengths. In both cases we see an exponential decay with energy, however for a value close to resonance $a_s= -0.315 a_0$ (blue line) the overlaps tend to be one or two orders of magnitude larger than they are for an interaction strength far from resonance $a_s= -0.2a_0$ (red line). This underscores the importance of considering a comprehensive continuum description, especially in proximity to a resonance.

To calculate the bound states, we use a basis of around 20000 B-splines spanning the entire box and standard diagonalization routines designed for sparse matrices. This step is especially important in order to accurately obtain bound states close to threshold which decay very slowly at large $r$.

\subsection{Ion in the high density limit}
Here we show the absorption spectrum of an ionic impurity for the same parameters in the text, except at a density ten times greater. 
Here, the Gaussian lineshape is again clear, and the peak position follows $E_\mathrm{Ion}$. Some deviation from the smooth Gaussian can be seen near to one of the Ramsauer Townsend zeros.

	\begin{figure}[t]
		\includegraphics[width=1 \columnwidth]{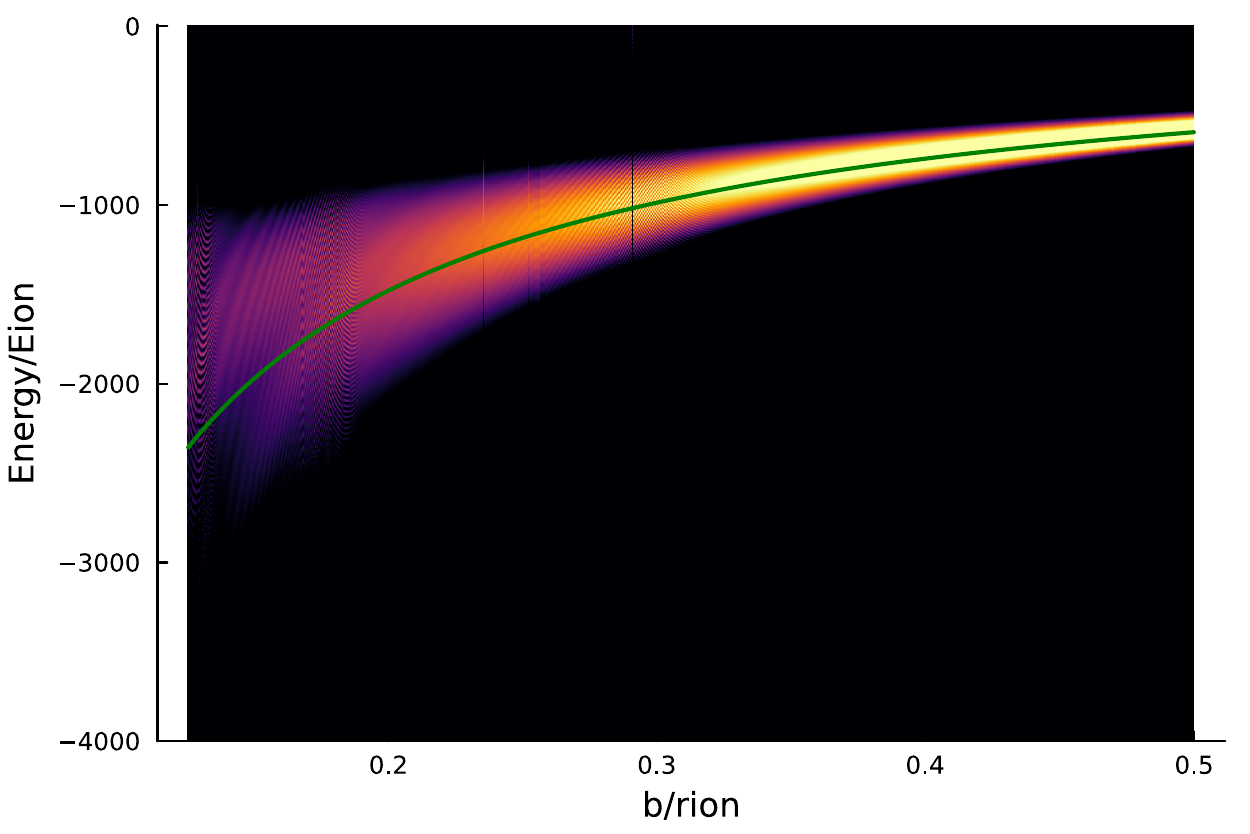}
		\caption{$A(\omega)$ of an ionic impurity at $\rho = 30\,R_\mathrm{Ion}^{-3}$. The green line shows the mean-field energy shift $E_\mathrm{Ion}$.  }\label{fig:iondens}
	\end{figure}


\end{document}